# A Hybrid Approach for Co-Channel Speech Segregation based on CASA, HMM Multipitch Tracking, and Medium Frame Harmonic Model

Ashraf M. Mohy Eldin
Department of Computer Engineering
Arab Academy for Science, Technology and Maritime Transport
Cairo, Egypt

Aliaa A. A. Youssif
Department of Computer Science
Faculty of Computers and Information
Helwan University
Cairo, Egypt

*Abstract*—This paper proposes a hybrid approach for co-channel speech segregation. HMM (hidden Markov model) is used to track the pitches of 2 talkers. The resulting pitch tracks are then enriched with the prominent pitch. The enriched tracks are correctly grouped using pitch continuity. Medium frame harmonics are used to extract the second pitch for frames with only one pitch deduced using the previous steps. Finally, the pitch tracks are input to CASA (computational auditory scene analysis) to segregate the mixed speech. The center frequency range of the gamma tone filter banks is maximized to reduce the overlap between the channels filtered for better segregation. Experiments were conducted using this hybrid approach on the speech separation challenge database and compared to the single (non-hybrid) approaches, i.e. signal processing and CASA. Results show that using the hybrid approach outperforms the single approaches.

## I. INTRODUCTION

In everyday life, speech doesn't arrive to our ears in a clean way, but rather corrupted by various types of noise including speech of other competing talkers in what is known as cocktail party effect. Human auditory system is remarkably capable of focusing on the target speech and separating it from noise.

On the contrary, artificial speech processing systems are designed to deal with clean, noise free speech. These systems need a front end component that segregates the target speech from other interferences. Competing speech is the most difficult kind of interference because of the similarity of temporal and spectral characteristics between target and interfering speeches. Work on speech segregation dates back to 70s [1].

The complexity of the speech segregation problem is related to the number of speakers and channels (i.e. microphones) used to record the speeches. The most difficult situation is when only one channel is used, i.e., co-channel speech segregation, as all spatial cues used in segregation are lost, e.g., inter-aural time, phase, and level differences.

Many approaches have been investigated to solve the co-channel speech segregation problem. The earliest approachesare the general signal processing approaches. R. H. Frazier [3] used adaptive comb filter with frequency spacing of pass bands varying with the fundamental frequency of the speech. T. F. Quatieri and R. G. Danisewicz [4] used the sinusoidal model of speech which assumes that speech consists of a sum of sin waves with varying amplitudes, frequencies, and phases over time. They used a minimum mean-squared error estimation combined with the sinusoidal model. D. S. Benincasa and M. I. Savic [5] used a technique to separate the co-channel mixed speech of 2 talkers by using constrained nonlinear optimization to separate overlapping voiced speech.

Another category of approaches is the computational auditory scene analysis (CASA). G. J. Brown and D. L. Wang [7] explained how CASA could be used in speech segregation showing how both monaural and binaural cues could be used for co and multichannels speech segregation. They also explained how to integrate CASA with speech recognition. L. Ottaviani and D. Rocchesso [9] proposed a system with 2 stages, pitch analysis using enhanced summary autocorrelation function (ESACF) and signal re-synthesis using highly zero-padded Fourier transform and its inverse. P. Li *et al.* [18] used objective quality assessment of speech (OQAS) combined with CASA. They used OQAS as a guide to lead CASA grouping. X. Zhang *et al.* [19] introduced the new concept of dynamic harmonic function (DHF) and replaced the conventional autocorrelation function (ACF) with DHF to suppress invalid peaks. Blind source separation (BSS) is a statistical approach that tries to recover a set of original signals from observed mixtures by assuming the linearity of the mixing process. A standard approach of BSS is independent component analysis (ICA) [10], which assumes the statistical independency of all sources. ICA needs 2 conditions to be satisfied to solve the speech segregation problem, namely, there must be a number of observed mixtures equal to or greater than the number of source signals, and all source signals must be perfectly aligned [2].

Obviously, these 2 conditions are not met in co-channel speech segregation problem. To solve this problem, a so called underdetermined BSS was invented [11]. In this technique, a priori knowledge obtained through training must be available. An example of the use of underdetermined BSS can be found in [12], in which case the priori knowledge was a set of time domain basis functions learned in a training phase. A comparison between CASA and BSS can be found in [13]. The comparison yields that CASA is more suitable to natural situations as it does not need a lot of conditions required by BSS. However, in the presence of these conditions, BSS may outperform CASA. This may suggest the join of the 2 approaches in a hybrid one.

Model based approaches could be used to solve the problem of co-channel speech segregation. These approaches consist of 3 steps. First, training phase is used to obtain patterns of sources. Second, patterns whose combinations model the observed signal are chosen. Third, selected patterns are used to estimate the sources directly or used to build filters to get the sources from the filtered observed signals. A. M. Reddy and B. Raj [14] used a model with minimum mean squared error estimator for co-channel speech segregation. H. A. T. Kristjansson and J. Hershey [15] used the male and female speech fine structure and the source signals strong high frequency resolution model. S. T. Roweis [16] used a simple factorial hidden Markov model (HMM) system which is trained on recordings of single speakers and then uses the co-channel observed signal to separate the mixture by calculating the masking function and re-filtering. The masking function is simply a non-stationary reweighting of the individual speakers' sub bands. D. E. M. J. Reyes-Gomez and N. Jojic [17] broke the mixed speech signal into multiple frequency bands. For each individual band they built separate HMM. Those separate HMMs are coupled together to model the whole mixed speech.

Hybrid approaches try to benefit the advantages of 2 or more different approaches to get better segregation results. J. Ming et al. [25] combined a missing feature technique to improve the robustness against crosstalk and noise with: Wiener filter to enhance the speech, hidden Markov model to reconstruct the speech, and speaker dependent/independent modeling to recognize both the speaker and the speech. P. Li et al. [26] combined Gaussian mixture models (GMMs) and max vector quantizers (MAXVQ) with CASA to separate co-channel mixed speech. Pitch is considered the most important cue in CASA as it is used for both simultaneous and sequential grouping. Accordingly, multipitch determination algorithms (MPDAs) were the subject of many researches. MPDAs may be either: time-domain, frequency domain, or time-frequency-domain. Time-domain MPDAs depend on the speech waveform temporal characteristics, e.g., autocorrelations. Frequency-domain MPDAs uses the short term spectrum to detect the fundamental frequency. Time-frequency-domain MPDAs obtain the signals using multichannel front end, then band-filter these signals, and finally perform time-domain analysis [2].

An example of time domain MPDAs is presented byA. de Cheveigne[22]. He extended the average magnitude difference function (AMDF) in a two-dimensional way by cancelling one of the 2 speakers and estimating the pitch of the other one. An example of frequency domain MPDAs is the one suggested by F. Sha and L. K. Saul [23]. They proposed an approach with instantaneous frequency estimation as front end and nonnegative matrix factorization as back end. An example of time frequency MPDAs is the one introduced by M. Wu et al. [24]. In their approach, periodicity information is extracted across different frequency channels, and the pitch tracks are formed using HMM.

The proposed approach combines a model based MPDA with time and frequency one for multipitch tracking. The enhanced pitch tracks are used as cues for enhanced CASA segregation.

## II. PROPOSED APPROACH OVERVIEW

The proposed approach consists of 2 stages of enhancements, pitch tracking enhancement and CASA segregation enhancement, Fig. 1. The 1st stage consists of 4 steps, HMM multipitch tracking, prominent pitch enrichment, grouping based on pitch continuity, and extracting the 2nd pitch (for frames with only one pitch deduced) using medium frame harmonics.

The chosen algorithm for multipitch tracking is the one developed by Z. Jin and D. L. Wang [21]. The reason behind choosing this algorithm is the ability to use it for different speech corpuses without the need to be trained on the new ones. This makes it a general algorithm and more preferable than other algorithms. Although it was developed taking reverberation into account, it is usable for normal conditions.

Medium frame harmonic extraction is inspired by the work of Q. Huang and D. Wang [6]. They used both short and long frames for pitch state deduction and pitch calculation. However, in the proposed approach, pitch hypothesis from the MPDA of Z. Jin and D. L. Wang [21] besides the concept of pitch continuity are used to judge the pitch state. So, only one type of frames is used and this type was neither short nor long but medium to enhance the resolution of the Fourier analysis without compromising neither the stationary assumption needed for Fourier transform nor the calculation speed.

For the segregation stage CASA algorithm proposed by G. Hu and D. L. Wang [8] is used with some changes. The center frequency range of the gamma tone filter banks was maximized to reduce the overlap between the channels filtered for better segregation. Also a new mechanism for time frequency (T-F) unit labeling that depends on the pitch tracks of both talkers was used.

## III. HMM MULTIPITCH TRACKING

The list describes the changes from the base algorithm:

- The base algorithm assumed that the sampling frequency of the mixed speech is 16 KHz. For this paper, the sampling frequency of the test database is 25 KHz. This means that the typical pitch range for both male and female from 80 to 500 Hz or time lags from 2 to 12.5 msnow corresponds to 50 to 313 samples, i.e. pitch state space which consists of the union of 0, 1, and 2 pitch hypothesis, since this algorithm tracks up to 2 pitches simultaneously,is described by (1).

$$S = S_0 \cup S_1 \cup S_2 \qquad (1)$$

Where $S_0 = \{\phi\}$, $S_1 = \{\{\tau_1\} : \tau_1 \in [50,313]\}$, and $S_2 = \{\{\tau_1,\tau_2\} : \tau_1,\tau_2 \in [50,313], \tau_1 \neq \tau_2\}$. $\phi$ indicates the absence of a pitch, and $\tau_1$ and $\tau_2$ represent the time lags of the 2 pitches in sample points.

- Z. Jin and D. L. Wang mentioned that broadband noise distorts the spectral peaks of the speech and causes

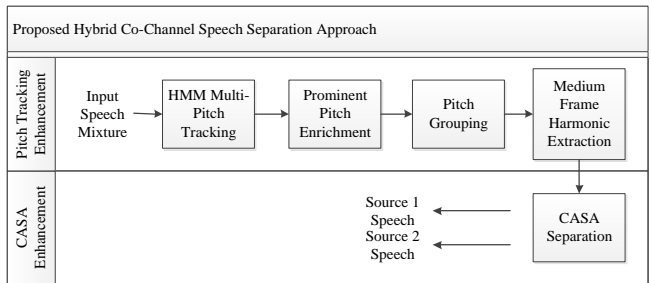

Fig. 1 Proposed Hybrid Approach

HMM search to be biased towards $S_2$. To overcome this, they performed two independent Viterbi searches. The 1st assumes the presence of one pitch in the frame at maximum. The 2nd performs the search normally trying to find up to 2 pitches in the frame. They used linear discriminant analysis (LDA) to decide which one of the 2 searches is correct for the frame in question.

Since the database used to test the proposed approach is noise free, this step is removed and the search is always performed for up to 2 pitches in the frame.

This enhanced the results, although slightly, however it speeded up the whole process in a good way.

### IV. PROMINENT PITCH ENRICHMENT

HMM multipitch tracks are enriched with the prominent pitch calculated using summary autocorrelation across all channels. For each frame, if zero pitch is deduced by the HMM multipitch tracking, i.e. zero pitch hypothesis, $S_0$, this hypothesis is unconditionally overridden to one pitch hypothesis, $S_1$ and the pitch is simply the prominent pitch. If one pitch is deduced by the HMM multipitch tracking, i.e. one pitch hypothesis, $S_1$, if the prominent pitch is far from the pitch deduced by more than a threshold, 32, the one pitch hypothesis is overridden to two pitch hypothesis, $S_2$, and the second pitch is simply the prominent pitch.

### V. PITCH GROUPING

Pitches are assigned to the proper source, 1 and 2, based on pitch continuity as follows:

- The previous frame in the following paragraph is the first frame preceding the current one with more than 0 pitches deduced.

- If the current and previous frames both have 2 pitches deduced, then the pitches of the current frame are assigned to the tracks that achieves the minimum distance between corresponding pitch lags.

- If the current frame has 2 pitches and the previous frame has only one pitch deduced, then the current frame pitch nearest to the previous frame one is put in the same track and the other current frame pitch is put in the opposite track.

- If the current frame has one pitch and the previous frame has 2 pitches deduced, then the current frame pitch is put in the same track of the nearest pitch of the previous frame.

- If the current and previous frame has only one pitch deduced, then if the distance between the 2 pitch lags is within a threshold, 32, then the current frame pitch is put in the same track as the previous frame pitch, otherwise, in the opposite track.

### VI. MEDIUM FRAME HARMONIC EXTRACTION

The following steps are used to get the other source's pitch, for frames with only 1 pitch obtained:

- Previous pitch, $F_{prev}$ for the track that needs to estimate the current pitch is obtained by iterating previous frames the same way done in pitch grouping.

- Also, next pitch, $F_{next}$ is obtained by iterating next frames.

- Fourier transform is obtained for the current frame with medium frame length, 50 ms that is sufficient to get good resolution of the harmonics compared to short frame of 30 ms while maintaining less complexity of calculation compared to long frame of 90 ms.

- All Fourier components after a threshold of 4000 Hz are removed. This is because the main energy of voiced speech is concentrated in the low frequency.

- The remaining Fourier components are divided into bands of 200 Hz. For each band peaks that are not less in magnitude than 1/5 of the highest peak in the band are obtained. They will form a vector of chosen harmonics for both pitches of the 2 sources, $F_{vec}$.

- From $F_{vec}$, all the harmonics (multiple integers and also half, quarter, and 1/8) that belongs to the pitch already known for that frame, F, including the pitch itself, are removed. Now, the vector contains only candidates of the other pitch and their harmonics.

- Candidate pitches, $F_{cand}$ are those that exist in Fvec and are not far from $F_{prev}$ or $F_{next}$ by 8 Hz.

- For each pitch in $F_{cand}$, the harmonic order (number of harmonics that exist in $F_{vec}$) and the average frequency deviation of those harmonics from the ideal ones (multiple integers of the candidate pitch) is calculated. Only those pitches with harmonic order that is not less than 9/10 of the maximum harmonic order are chosen. The pitch with minimum frequency deviation from the chosen ones is simply the other pitch for that frame.

### VII. CASA SEGREGATION

The list describes the changes from the base algorithm:

- Before segregation, pitch tracks need to be refined from pitches suspected to be error. This enhances the segregation as leaving a T-F unit without assigning to a source (to be assigned later based on grouping) is better than assigning it to the wrong source. Pitches in tracks less than 5 contiguous frames are considered suspected and removed.

- In the peripheral analysis, input signal is passed through a bank of 128 gamma tone filters centered

from 21 to 12500 Hz instead of from 80 to 5000 Hz. The frequency range is maximized to reduce the overlapping between gamma-tone filtered channels for better segregation.

- Unit Labeling:

Both pitch tracks are used to label each T-F unit as either belonging to source 1 or 2 instead of just using the target pitch track to label the T-F unit as either belonging to the target source or interference. The following points worth mentioning:

- Error in segregation is directly related to error in pitch tracking.

- When using only the pitch track of the target source, error in separating target source is directly related to error in the pitch track of the target source.

- When using pitch tracks of both users, error in segregation of both sources is directly related to the average error of both pitch tracks.

- Accordingly, if the interest is to only separate one source (target source), if the target pitch track error is less than the average error, then using only target pitch track in unit labeling is better than using both pitch tracks. Otherwise, using both pitch tracks is better.

- However, if the interest is to separate both sources, it is always better to use both pitch tracks.

Features used in unit labeling are correlogram $A_H$ (2) and envelope correlogram $A_E$ (3).

$$A_H(c,m,\tau) = \frac{\sum_n h(c,mT_f - nT_s)h(c,mT_f - nT_s - \tau T_s)}{\sum_n h^2(c,mT_f - nT_s)} \quad (2)$$

$$A_E(c,m,\tau) = \frac{\sum_n h_E(c,mT_f - nT_s)h_E(c,mT_f - nT_s - \tau T_s)}{\sum_n h_E^2(c,mT_f - nT_s)} \quad (3)$$

Where $c$ is the channel number, $m$ is the frame number, $\tau$ is the time delay at which autocorrelation $A$ is calculated, $n$ is the digitized time, $T_f$ is the time shift from one frame to the next (10 ms) and $T_s$ is the sampling time, and $h$ is the channels response transduced by Meddis model.

Unit labeling is done using the following steps:

- If the current frame has 0 pitch deduced, then all channels T-F units are not labeled.

- If the current frame has only 1 pitch deduced, if AH/max(AH)>0.85 if the unit is marked as 1, or AE/max(AE)>0.7 if the unit is marked as 2, the unit is labeled to the same track of the pitch, otherwise, it is labeled to the opposite track.

- If the current frame has 2 pitches deduced, the previous ratios are calculated for both pitches. The unit is labeled to the track of the pitch that satisfies the condition.

However, if neither or both pitches satisfy the condition, the unit is not labeled at all.

VIII. EXPERIMENTAL RESULTS

*A. Database*

A different database was chosen to test the proposed approach rather than using the database used to test the base algorithms.

This is to make sure that these base algorithms (as was mentioned by their authors) are easily generalized to new speech corpuses with no training needs. The speech separation challenge [27] was particularly chosen because it is considered one of the most complex databases used in co-channel speech segregation problem. This is because it uses a small vocabulary set which leads to close similarity of the speeches of the competing talkers which makes the pitch tracking and segregation more difficult. In order to get accurate results, the whole database, a total of 900 mixed speech samples, was used rather than selecting a subset. The test was conducted on 0dB target to masker ratio (TMR) which is considered the most difficult situation in co-channel speech segregation problem as both talkers equally masks each other.

*B. Pitch Tracking Enhancements*

The pitch tracking enhancement stage will be compared with the base HMM multipitch tracking algorithm [21]. The comparison measures will be the gross error, $E_{gs}$, and the fine error, $E_{fn}$. $E_{gs}$ is the percentage of frames where the deduced pitch differs from the ground truth pitch by more than 20%. $E_{fn}$ is the average deviation from the ground truth pitch for the frames with no gross error. Ground truth pitches are calculated using summary autocorrelation of the frames of the original speeches of each source before mixing. This is not error free, some ground truth pitches will not be correct. However, the same error will be added (approximately) to both the base approach and the proposed one and will not affect the comparison. Since this paper is interested in separating the speech of both sources, the sum of gross errors of both pitch tracks and the sum of fine errors using the enhanced proposed pitch tracking stage will be compared with the base algorithm. Table. I shows the pitch tracking results.

*C. Whole Segregation Approach Enhancements*

The proposed approach will be compared to: the one proposed by G. Hu and D. L. Wang [8], the one proposed by Z. Jin and D. L. Wang [20], and the traditional harmonic selection approach. The following points need to be stated:

- The multi pitch tracker proposed by M. Wu *et al.* [24] and used in [8] is replaced by the base pitch tracker [21] since [24] needs to be trained on the speech separation challenge and the target of this paper is to enhance algorithms that need no training.

- The same suggested unit labeling step that uses both pitch tracks will be used for the base algorithm [8].

- The previous 2 points actually enhances the algorithm [8]. If the comparison of the proposed approach was done with [8] without these 2 points, much better

TABLE 1. PITCH TRACKING RESULTS

|  |  | Average $E_{gs}$ % | | Average $E_{fn}$ % | |
| --- | --- | --- | --- | --- | --- |
|  |  | Base Approach | Proposed Approach | Base Approach | Proposed Approach |
| Different Gender | | 59 | 51.9 | 4 | 4.9 |
| Same Gender | Male | 52.7 | 46.7 | 4.8 | 5.5 |
|  | Female | 68.8 | 59.2 | 5 | 5.8 |
| Same Talker | Male | 55.6 | 48.1 | 5.4 | 5.6 |
|  | Female | 69.1 | 59.1 | 5.1 | 5.5 |

results are obtained. However, this is done to get more accurate comparison results.

- Since the test database contains only speech interference, the speech model of [20] is used only for faster calculations. The binary masks for both talkers are used to label the T-F units as opposed to [20] which only uses the target talker's mask. T-F unit is labeled to talker 1 if mask 1 = 1 and mask 2 = 0. Similarly, T-F unit is labeled to talker 2 if mask 1 = 0 and mask 2 = 1. In the grouping stage, only T-F units labeled were taken into consideration.

- Harmonic selection is included in the comparison to show that CASA approaches outperform the traditional signal processing approaches. The pitch tracks obtained using the base pitch tracker [21] was used to select the harmonics of both talkers (the proposed enhanced pitch tracker stage could also be used for comparison and the same conclusion of the superiority of CASA approaches to the signal processing ones would be reached). If 2 pitches are tracked for a frame, those 2 pitches are used to select the harmonics of both talkers. If only one pitch is tracked, it is used to select the harmonics of the respected talker. The other talker's signal is the difference between the mixed signal and the first talker's one. If no pitch is tracked, the choice is to assign the frame to the first talker.

The comparison measure is the signal to noise ratio (SNR). CASA segregation gives some amplification (though not uniform) to the segregated speech. In order for the SNR to be accurate, the original speech had to be compensated for such amplification. This was accomplished by applying all the stages of CASA to the original speech except using all 1s mask for unit labeling. Original speech was left as it is for SNR calculation of harmonic selection. Since this paper is interested in both talkers' speeches, the average SNR of both talkers will be compared. Table. II shows the segregation results.

IX. ANALYSIS AND DISCUSSION

*A. Pitch Tracking Enhancements*

The proposed pitch tracking enhancement stage shows better results than the base approach. Enhancements range from 6 to 10% decrease in $E_{gs}$ with overall average of about 8%. As expected, the decrease in $E_{gs}$ results in increase in $E_{fn}$ since more pitches now are taken into account while calculating $E_{fn}$ with their deviation from the ground truth pitch added. However, the increase in $E_{fn}$ is in the range of 0.2 to 0.9% with overall average of 0.6%. This is a small increase in $E_{fn}$ which suggests that the pitches obtained from the enhancement stage are very close to the ground truth pitches with small deviation.

The following points are worth mentioning:

- $E_{gs}$, whether for base and proposed approaches, is very high. This is due to the nature of the speech separation challenge database as explained in sec. VIII-A. It is expected to have better (less) $E_{gs}$ for normal speech conditions with less similarity between competing talkers speeches.

- The base algorithm has a problem in tracking the female pitches. This is suggested from the higher $E_{gs}$ for female same gender and same talker mixtures than corresponding male mixtures. $E_{gs}$ for same gender female mixture is higher than male mixture by 16.1%. For same talker, it is higher by 13.5%. The proposed approach enhanced same gender female pitch tracking by 9.6% as opposed to 6% for male. Also, for same talker, the proposed approach enhanced female tracking by 10% as opposed to 7.5% for male. This means that the proposed approach could deduce some of the female pitches missed by the base approach, mostly in the prominent pitch enrichment step.

*B. Whole Segregation Approach Enhancements*

The proposed approach shows better segregation results than the compared approaches. Enhancements range from 3.1% for the different gender case to 27% for same talker female case. The following points are worth mentioning:

- Enhancements are better when the case is worse. This is apparent from the fact that less enhancements happened in different gender case whereas best ones happened in same talker case. Also, enhancements for male-male are less than enhancements for female-female cases. The following points will try to explain the reason behind this.

- Different gender case exhibits minor enhancements. This is due to that the proposed approach depends on enhancing the pitch tracking and minimizing the T-F units overlapping between the 2 talkers. For the different gender case, the overlapping was originally small because female T-F units tend to be in the higher frequency channels whereas male T-F units tend to be in the lower frequency channels.

- For same gender and same talker cases, there was a reasonable amount of overlapping between the T-F units of both talkers as their T-F units tend to occupy the same range of frequencies. This means that the step of minimizing the overlap made better enhancements in these cases than the different gender case.

- This also means that the step of pitch tracking enhancements made fewer enhancements than the step of overlap minimization. This suggests that more enhancements in pitch tracking are needed.

TABLE II. SEGREGATION RESULTS

| | | SNR dB | | | |
|---|---|---|---|---|---|
| | | G. Hu and D. L. Wang [8] | Proposed Approach | Harmonic Selection | Z. Jin and D. L. Wang [20] |
| Different Gender | | 3.2 | 3.35 | 0.83 | 3.25 |
| Same Gender | Male | 2.44 | 2.83 | 0.6 | 2.43 |
| | Female | 2.2 | 2.68 | 1.02 | 2.21 |
| Same Talker | Male | 1.87 | 2.31 | 0.53 | 1.86 |
| | Female | 1.74 | 2.16 | 0.78 | 1.7 |

- The minor differences in enhancements between same gender and same talker cases and also between male and female in each case are due to the minor better enhancements in pitch tracking for these cases.

## X. CONCLUSION AND FUTURE WORK

This paper proposed enhanced co-channel speech segregation approach. It also proposed enhanced MPDA that could be used on its own for multiple purposes. More future enhancements are suggested in the following lists:

Enhancing the pitch tracking for female talkersby detecting their presence in the mix using their higher pitch range, then using different parameters for the base algorithm tuned for females than those tuned for males.

Using hybrid algorithm for segregation by detecting the frame state. For voiced-voiced or unvoiced-unvoiced, CASA approaches may be used for segregation. For voiced-unvoiced, low pass and high pass filters could be used to get the speech of each talker.

### ACKNOWLEDGMENT

The authors would like to thank G. Hu and D. L. Wang [8] and Z. Jin and D. L. Wang [20] [21], for providing their algorithms and codes.

### REFERENCES

[1] K. S. Ananthakrishnan and K. Dogancay(2009), "Recent trends and challenges in speech separation systems research - a tutorial review," in TENCON 2009, IEEE Region 10 Conference, Hong Kong, 23-26 Nov., 2009(C1).

[2] Y. Mahgoub, B.Eng., and M.Eng., " Co-channel speech separation using state-space reconstruction and sinusoidal modeling," Ph.D. dissertation, Carleton University, Ottawa, Ont., Canada, Canada, 2010.

[3] R. H. Frazier, "An adaptive filtering approach toward speech enhancement," M.S. thesis, Department of Electrical Engineering and Computer Science, Massachusetts Institute of Technology, Cambridge, MA, USA, June 1975.

[4] T. F. Quatieri and R. G. Danisewicz, "An approach to co-channel talker interference suppression using a sinusoidal model for speech," IEEE Transactions on Acoustics, Speech, and Signal Processing, vol. 38, pp. 56-69, Jan. 1990.

[5] D. S. Benincasa and M. I. Savic, "Co-channel speaker separation using constrained nonlinear optimization," in ICASSP-1991, Munich, Germany, Apr. 1997, vol. 2, pp. 1195-1198.

[6] Q. Huang and D. Wang, "Single-channel speech separation based on long-short frame associated harmonic model," Digital Signal Processing Journal, vol. 21, issue 4, pp. 497-507, July 2011.

[7] G. J. Brown and D. L. Wang, "Separation of speech by computational auditory scene analysis," in Speech Enhancement (J. Benesty, S. Makino, and J. Chen, eds.), New York: Springer, 2005, ch. 16, pp. 371-402.

[8] G. Hu and D. L. Wang, "An auditory scene analysis approach to monaural speech segregation," in Topics in Acoustic Echo and Noise Control (E. Hansler and G. Schmidt, eds.), New York, NY, USA: Springer, 2006, ch. 12, pp. 485-515.

[9] L. Ottaviani and D. Rocchesso, "Separation of speech signal from complex auditory scenes," in COST G-6 Conference on Digital Audio Effects, Limerick, Ireland, Dec. 2001, pp. 87-90.

[10] A. Hyvarinen and E. Oja, "Independent component analysis: algorithms and applications," Neural Networks, vol. 13, pp. 411-430, May-June 2000.

[11] P. Bofill and M. Zibulevsky, "Underdetermined blind source separation using sparse representations," Signal Processing 81, 2001, pp. 2353–2362.

[12] G. Jang, T. Lee, and Y. Oh, "Single-channel signal separation using time-domain basis functions," IEEE Signal Processing Letters, vol. 10, pp. 168-171, June 2003.

[13] A. J. W. van der Kouwe, D. Wang, and G. J. Brown, "A comparison of auditory and blind separation techniques for speech segregation," IEEE Transactions on Speech and Audio Processing, vol. 9, pp. 189-195, Mar. 2001.

[14] A. M. Reddy and B. Raj, "A minimum mean squared error estimator for single channel speaker separation," in INTERSPEECH 2004 - ICSLP, Jeju Island, Korea, Oct. 2004, pp. 2445-2448.

[15] H. A. T. Kristjansson and J. Hershey, "Single microphone source separation using high resolution signal reconstruction," in ICASSP-2004, Montreal, QC, Canada, May 2004, vol. 2, pp. 817-820.

[16] S. T. Roweis, "One microphone source separation," in Advances in Neural Information Processing Systems (NIPS-2001) (T. K. Leen, T. G. Dietterich, and V. Tresp, eds.), Cambridge, MA, USA: MIT Press, Dec. 2001, vol. 13, pp. 793-799.

[17] D. E. M. J. Reyes-Gomez and N. Jojic, "Multiband audio modeling for single channel acoustic source separation," in ICASSP-2004, Montreal, Canada, May 2004, vol. 5, pp. 641-644.

[18] P. Li, Y. Guan, B. Xu, and W. Liu, "Monaural speech separation based on computational auditory scene analysis and objective quality assessment of speech," IEEE Transactions on Audio, Speech, And Language Processing, vol. 14, no. 6, Nov. 2006.

[19] X. Zhang, W. Liu, and B. Xu, "Monaural voiced speech segregation based on dynamic harmonic function," EURASIP Journal on Audio, Speech, and Music Processing vol. 2010, Article ID 252374, 2010.

[20] Z. Jin and D. L. Wang, "Reverberant speech segregation based on multipitch tracking and classification," IEEE Transactions on Audio, Speech, And Language Processing, vol. 19, no. 8, Nov. 2011.

[21] Z. Jin and D. L. Wang, "HMM-based multipitch tracking for noisy and reverberant speech," IEEE Transactions on Audio, Speech, And Language Processing, vol. 19, no. 5, July 2011.

[22] A. de Cheveigne, "A mixed speech F0 estimation algorithm," in EUROSPEECH-1991, Genova, Italy, Sep. 1991, pp. 445-448.

[23] F. Sha and L. K. Saul, "Real-time pitch determination of one or more voices by nonnegative matrix factorization," in Advances in Neural Information Processing Systems (NIPS 2004) (L. K. Saul, Y. Weiss, and L. Bottou, eds.), Cambridge, MA, USA: MIT Press, Dec. 2004, vol. 17, pp. 1233-1240.

[24] M. Wu, D. Wang, and G. Brown, "A multipitch tracking algorithm for noisy speech," IEEE Transactions on Speech and Audio Processing, vol. 11, pp. 229-241, May 2003.

[25] J. Ming, T. J. Hazen, and J. R. Glass, "Combining missing-feature theory, speech enhancement, and speaker-dependent/-independent modeling for speech separation," Computer Speech and Language vol. 24, pp. 67–76, 2010.

[26] P. Li, Y. Guan, S. Wang, B. Xua, and W. Liu, "Monaural speech separation based on MAXVQ and CASA for robust speech recognition," Computer Speech and Language, vol. 24, pp. 30–44, 2010.

[27] M. Cooke, J. R. Hershey, and S. J. Rennie, "Monaural speech separation and recognition challenge," Computer Speech and Language, vol. 24, pp. 1–15, 2010.